\documentclass{article}

\usepackage{arxiv}

\usepackage[utf8]{inputenc} % allow utf-8 input
\usepackage[T1]{fontenc}    % use 8-bit T1 fonts
\usepackage{hyperref}       % hyperlinks
\usepackage{url}            % simple URL typesetting
\usepackage{booktabs}       % professional-quality tables
\usepackage{amsfonts}       % blackboard math symbols
\usepackage{nicefrac}       % compact symbols for 1/2, etc.
\usepackage{microtype}      % microtypography
\usepackage{graphicx}
\usepackage[numbers]{natbib}
\usepackage{doi}
\usepackage{arydshln}
\usepackage{algorithm}
\usepackage{algpseudocode}
\usepackage[dvipsnames]{xcolor}

\title{A drug classification pipeline for Medicaid claims using RxNorm}

\author{Nicholas Williams \\
	Department of Epidemiology\\
	Columbia University\\
	New York, NY 10032 \\
	\texttt{ntw2117@cumc.columbia.edu} \\
        \And
        Kara E. Rudolph \\
	Department of Epidemiology\\
	Columbia University\\
	New York, NY 10032 \\
	\texttt{kr2854@cumc.columbia.edu} \\
}

\renewcommand{\headeright}{Technical Report}
\renewcommand{\shorttitle}{Classifying Drugs with RxNorm}

\begin{document}
\maketitle

\begin{abstract}

\noindent \textbf{Objective:} Freely preprocess drug codes recorded in electronic health records and insurance claims to drug classes that may then be used in biomedical research.
\\

\noindent \textbf{Materials and Methods:} We developed a drug classification pipeline for linking National Drug Codes to the World Health Organization Anatomical Therapeutic Chemical classification. To implement our solution, we created an \texttt{R} package interface to the National Library of Medicine's RxNorm API. \\

\noindent \textbf{Results:} Using the classification pipeline, 59.4\% of all unique NDC were linked to an ATC, resulting in 95.5\% of all claims being successfully linked to a drug classification. We identified 12,004 unique NDC codes that were classified as being an opioid or non-opioid prescription for treating pain.\\

\noindent \textbf{Discussion:} Our proposed pipeline performed similarly well to other NDC classification routines using commercial databases. A check of a small, random sample of non-active NDC found the pipeline to be accurate for classifying these codes. \\
 
\noindent \textbf{Conclusion:} The RxNorm NDC classification pipeline is a practical and reliable tool for categorizing drugs in large-scale administrative claims data.\\

\end{abstract}


\section{Background and Significance}

Electronic health record (EHR) and administrative insurance claims data have become an increasingly important asset in conducting observational epidemiological and pharmacological research \cite{casey2016using,farjo2024comparison}. Much of this research requires classifying drugs in terms of mechanism of action, therapeutic intent, or chemical structure; commonly referred to as \textit{drug class}. However, drug data in the EHR are typically recorded using codes that do not indicate the general drug class, and so require preprocessing before they can be used for biomedical research. In the United States (US), EHR drug data are recorded as national drug codes (NDCs), which are unique three-segment, 11-digit identifiers assigned to all human drugs. The first segment is a manufacturer identifier, the second segment is a product identifier, and the third is a package identifier. Although an NDC uniquely identifies a specific drug product, it does not directly identify the drug class. Thus, preprocessing these data by converting NDC to useful study-specific drug classes is a key step in applied research. 

One option for classifying NDC is to pay for a commercial database\cite{homer2016drug}, such as IBM Micromedex Red Book \cite{micromedex} or Multum from Oracle Health \cite{multum}. However, paying for one of these services limits access to researchers with large enough budgets, and introduces a roadblock for reproducibility. Another approach would be to manually create a list of all the NDCs that belong to a given drug class of interest \cite{CDCOpioidsData,miller2019prevalence}. While free, this approach introduces variability between research teams, is time-consuming, and may have poor performance as the number of NDC that belong to a certain class may be well into the thousands (see Table \ref{tab:searchsummary}).

We address this gap by providing a freely available user-friendly software that performs this preprocessing step, linking NDCs with drug class codes. In performing this linkage, we use the free-to-use US National Library of Medicine's (NLM's) RxNorm, which links many products to different drug classification systems. We use RxNorm's linkage to the World Health Organization’s (WHO's) Anatomical Therapeutic Chemical classification (ATC), which classifies substances "according to the organ or system on which they act and their therapeutic, pharmacological and chemical properties." \citep{whoATC2023} In the ATC system, drugs are assigned a code that indicates a hierarchical subdivision into five different levels with increasing levels of specificity. We are only interested in the first four ATC levels as the fifth level indicates the specific substance. For example the ATC code, up to the fourth level, for fentanyl is N01AH: "N" indicates fentanyl operates on the nervous system, "01" indicates fentanyl is an anesthetic, the "A" indicates it's part of the "general" subgroup of anesthetics, and the "H" indicates it's an opioid anesthetic.

\section{Objective}

We were motivated to develop this software by the need to identify opioid and non-opioid drugs prescribed for pain in a large cohort of Medicaid claims data. We use our motivating example to demonstrate the software's utility and performance. Our software uses the NLM RxNorm API and an \texttt{R} package interface. We evaluate its performance by the proportion of prescription claims that are linked to a drug class. 

\section{Materials and methods}

\subsection{Classification pipeline}

Our approach retrieves drug information using a set of free-to-use application program interfaces (APIs) provided by NLM via representational state transfer (REST) to access the RxNorm and RxTerms datasets: the RxNorm and RxTerms \citep{RxNorm}, and RxClass \citep{RxClass} APIs. To avoid sending potentially thousands of requests to the NLM servers, we also make use of RxNav-in-a-Box, a locally installable version of the APIs, in combination with Docker \citep{merkel2014docker}. To query the APIs, we created an \texttt{R} \citep{R} package interface which is available for download on \href{https://github.com/nt-williams/rxnorm}{GitHub}. 

First, we retrieve an NDC's status. In RxNorm, an NDC is classified as either "active", "obsolete", "alien", or "unknown". 
We ignore NDCs with a status of "unknown" as they are items that have never appeared in RxNorm. Often these products are medical supplies, vitamins or minerals, dietary drugs, and other over-the-counter medications. 

If an NDC is considered "active", we then query RxNorm for the RxCUI assigned to the NDC and then submit a query for the associated product-level ATC. If a product-level ATC is not found, we instead submit a query for the ingredient-level ATC of all active-ingredients in the drug. Note that product-level ATC are assigned by NLM, while ingredient-level ATC are assigned to a substance by the WHO. 

If an NDC is considered "obsolete" or "alien", we again query RxNorm for the RxCUI assigned to the NDC, but then perform an additional query for the RxCUI status to see if the RxCUI has been "remapped", "quantified", or considered "not current" or "obsolete". If the RxCUI is "remapped" or "not current", we attempt to replace the original RxCUI with the current RxCUI, and again search for the product or ingredient level ATC. If the RxCUI status is "obsolete" we attempt to link the NDC to a new RxCUI through the \textit{semantic clinical drug} (SCD) concept which itself is a mapping of a drug product into its ingredient, strength, and dose form; again, then searching for the product or ingredient level ATC of the RxCUI associated with the SCD. If the RxCUI status is "quantified", it indicates that the RxCUI is missing a "quantity factor" but has been linked to another related concept that may be quantified, which we then again attempt to link to a product or ingredient-level ATC. 

The \texttt{R} script used for running the above classification pipeline is available on \href{https://github.com/CI-NYC/rxnorm-paper/tree/main}{GitHub}.

\begin{table}[h]
    \centering
    \caption{ATC codes of interest}
    \begin{tabular}{rp{8.5cm}c}
        \toprule
        ATC & Description & No. of NDC linked \\
        \midrule
        \multicolumn{2}{l}{\textit{Opioids for pain}} \\
        \addlinespace[0.2em]
        \hdashline
        \addlinespace[0.4em]
        N02A & Opioid analgesics & 2264\\
        N01AH & Opioid anesthetics & 117 \\
        N07BC & Drugs used in opioid dependence & 169\\
        \midrule
        \multicolumn{2}{l}{\textit{Non-opioids for pain}} \\
        \addlinespace[0.2em]
        \hdashline
        \addlinespace[0.4em]
        A03D & Antispasmodics in combination with analgesics & 0 \\
        A03EA & Antispasmodics, psycholeptics and analgesics in combination & 0\\
        M01 & Anti-inflammatory and anti-rheumatic products & 2529\\
        M02A & Topical products for joint and muscular pain & 295\\
        M03 & Muscle relaxants & 848 \\
        N02B & Other analgesics and anti-pyretics & 3154\\
        N06A & Anti-depressants & 2628\\
        \midrule
        \multicolumn{2}{r}{Total:} & 12004 \\
        \bottomrule
    \end{tabular}
    \label{tab:searchsummary}
\end{table}

\subsection{Data}

Our dataset includes prescription drug claims in the Medicaid T-MSIS Analytic Files (TAF) pharmacy and other services files from non-pregnant adults aged 35-64 years who were non-dual-eligible Medicaid beneficiaries enrolled 2016-2019 from the following 26 states that implemented Medicaid expansion under the Affordable Care Act in or prior to 2014: ND, VT, NH, CA, OR, MI, IA, NV, OH, IL, NY, MD, MA, RI, HI, WV, WA, KY, DE, AZ, NJ, MN, NM, CT, CO, AR \citep{KaiserMAP}. We limited our dataset to valid NDC that were 11-digits long and only contained values 0-9. We refer the reader to \citet{hoffman2023independent} for further background.

\section{Results}

Our goal was to identify opioids and non-opioids typically prescribed for treating pain. The classification pipeline was run on a 2021 MacBook Pro with an Apple M1 Pro chip and 32GB of RAM. Using parallel processing, the pipeline ran in approximately 12.5 minutes. The ATC codes used to identify these categories and the number of NDC that were found to belong to an ATC class are shown in Table \ref{tab:searchsummary}. In total, we identified 12,004 NDC codes that belonged to one of the opioid and non-opioid ATC categories of interest.

The dataset consisted of 809,484,945 claims and 126,604 unique NDC. Of the unique NDC, 50,142 (39.6\%) had an NDC status of "active"; 23,520 (18.6\%) were "obsolete"; 34,267 were "unknown" (27.1\%); 18,675 were "alien" (14.8\%). Among the "active" NDC, 49,861 (99.4\%) were successfully mapped to a product-level or ingredient-level ATC; 20,946 (89.1\%) of the "obsolete" NDC were mapped to an ATC; 4,484 of the "alien" NDC were linked to an ATC. Of all NDC, 75,270 (59.4\%) were successfully mapped to an ATC. However, the classified NDCs accounted for 95.5\% of all the non-unique NDCs in the dataset. 

Among the NDC that did have a linked RxCUI but did not have an ATC associated with the RxCUI 13,289 (77.4\%) had an RxCUI status of "not current", 1,974 (11.5\%) had an RxCUI status of "obsolete", 1,195 (7\%) had an RxCUI status of "remapped", and 75 (0.4\%) had an RxCUI status of "quantified". 635 "active" RxCUI were unsuccessfully mapped to an ATC. Of the "obsolete" RxCUI, 1,078 (54.6\%) were matched to an RxCUI with an associated ATC; 947 (79.2\%) of the "remapped" RxCUI and 74 (99\%) of the "quantified" RxCUI were matched to a new RxCUI with a corresponding ATC. As an audit for accuracy, we took a 0.01\% sample of the non-"active" NDC that were linked to an ATC code and did an ad-hoc search for their appropriate ATC category. The results of this check are shown in Table \ref{tab:audit}. There were two NDC incorrectly classified NDC; both NDC were classified as "R05D" (cough suppressants, excl. combinations with expectorants) while the correct classification was "R05F" (cough suppressants and expectorants, combinations).

Table \ref{tab:unmatched} shows the concept names for the five most common "active", "alien", or "obsolete" NDC and the five most common "unknown" NDC that were not linked to an ATC code. Of these, six were glucose testing strips, two were condoms, one was needle and one was a chamber used for an inhaler.

\begin{table}[h]
    \centering
    \caption{Random sample of "alien" or "obsolete" NDC that were linked to an ATC code and checked for ad-hoc accuracy.}
    \begin{tabular}{rccc}
        \toprule
        NDC & NDC Status & Pipeline ATC & Ad-hoc ATC \\
        \midrule
        00173024955 & Obsolete & 	     C01AA & C01AA \\
        11917014765 & Alien    & 	     A11CC & A11CC \\
        57770005500 & Obsolete & 	     S01XA & S01XA/S01KA \\
        00781286531 & Obsolete & 	     A02BA & A02BA \\
        66870050512 & Alien    &         A02AA & A02AA \\
        51079012220 & Obsolete &   A02BD/P01AB & P01AB \\
        51079078719 & Obsolete & 	     C10AB & C10AB \\
        00904526161 & Obsolete & 	     A02BA & A02BA \\
        11822074050 & Obsolete & 	     N02BE & N02BE \\
        50428322435 & Alien    & 	     D06AX & D06AX \\
        40986001765 & Alien    &         B03BA & B03BA \\
        43063005202 & Obsolete & 	     A04AA & A04AA \\
        43292055802 & Alien    &         A11GA & A11GA \\
        79854001535 & Alien    & 	     A11HA & A11HA \\
        00093213193 & Obsolete & 	     J01XE & J01XE \\
        60432004504 & Obsolete & 	     R05DA & R05FA \\
        52544093628 & Obsolete &   G03AA/G03AB & G03AB \\
        00440632530 & Obsolete & 	     C02AC & C02AC \\
        50428034138 & Obsolete & 	     N05CH & N05CH \\
        76439013004 & Obsolete & 	     M05BA & M05BA \\
        60687029825 & Obsolete & 	     N06BA & N06BA \\
        70030014843 & Alien    & 	     P03AC & P03AC \\
        50428689594 & Obsolete & 	     A02AF & A02AF \\
        54569317700 & Obsolete & 	     R05DA & R05FA \\
        45802091334 & Obsolete & 	     D10AE & D10AE \\
        \bottomrule
    \end{tabular}
    \label{tab:audit}
\end{table}

\begin{table}[h]
    \centering
    \caption{Examples of NDC and their concept names that were not matched to an ATC code.}
    \begin{tabular}{rp{8.5cm}}
        \toprule
        NDC & Concept Name \\
        \midrule
        \multicolumn{2}{l}{\textit{NDC with a status of "active", "obsolete", or "alien"}} \\
        \addlinespace[0.2em]
        \hdashline
        \addlinespace[0.4em]
        99073070822 & FREESTYLE LITE (GLUCOSE) TEST STRIP \\
        99073070827 & FREESTYLE LITE (GLUCOSE) TEST STRIP \\
        99073013001 & LANCET,FREESTYLE \\
        56151146004 & TRUE METRIX (GLUCOSE) TEST STRIP \\
        53885024450 & ONE TOUCH ULTRA (GLUCOSE) TEST STRIP \\
        \midrule
        \multicolumn{2}{l}{\textit{"unknown" NDC}} \\
        \addlinespace[0.2em]
        \hdashline
        \addlinespace[0.4em]
        94046000124 & ADVOCATE REDI-CODE TEST STRIP \\
        26893010002 & CONDOMS, LATEX, LUBRICATED \\
        08373917700 & PROCHAMBER HOLDING CHAMBER \\
        26893010102 & ATLAS CONDOM MIS LUB/COLR \\
        53885085401 & ONETOUCH KIT VERIO IQ\\
        \bottomrule
    \end{tabular}
    \label{tab:unmatched}
\end{table}

\section{Discussion}

EHR data are increasingly common and valuable source for health research. However, drug data in the EHR often require classification before being usable. Motivated by the need to classify opioid and non-opioid prescription claims for pain medications in Medicaid, we developed a free-to-use, open-source, and reliable drug classification pipeline using the NLM RxNorm API and \texttt{R}. 

Applying our proposed pipline to our motivating example of classifying prescription pain medications in Medicaid claims data, we were able to link 59.4\% of all valid unique NDC to an ATC code. Limiting to NDC with an NDC status of "active", "obsolete", or "alien", we linked 81.5\% of NDC to an ATC code. These results are similar to \citet{homer2016drug} who found that RxNorm linked 60\% of NDC to an ATC, \citet{peters2015approaches} who found that 84.2\% of NDC could be linked to a drug class when including historical versions of RxNorm, and \citet{ostropolets2024high} who linked 79.6\% of clinical drugs in RxNorm to ATC. The classified NDC accounted for 95.5\% of all the claims in the cohort which is on par with the 98.2\% coverage from \citet{homer2016drug} when using a commercial database. We evaluated the accuracy of the pipeline for classifying "obsolete" and "alien" NDC using a small random sample of the classified codes and found its performance to be reliable. 

A limitation of our pipeline is that we limited drug classification to ATC. RxNorm can link to other class sources, such as VA class or a classification for diseases that may be treatable with a drug product from the Medication Reference Terminology, both produced by the Veterans Health Administration. A future improvement to the pipeline could be to attempt to link the unclassified NDC to one of these other classes as a way to increase the claims coverage. While we used the pipeline to classify Medicaid claims, we expect it to be easily adaptable to other data as it only depends on NDC. Another limitation of approach is that it is limited to classifying drugs in the United States. Recently, however, \citet{reich2024ohdsi} developed RxNorm Extension which extends RxNorm to drugs outside of the United States. Future work could examine the performance of classifying drugs not used in the United States with ATC using this tool.

\section{Conclusion}

We developed a pipeline to link NDC to ATC in \texttt{R} using the National Library of Medicine's RxNorm API. Applying the classification pipeline to large-scale Medicaid claims dataset, we found it to be a practical and reliable tool for categorizing NDC.

\newpage
\bibliographystyle{unsrtnat}
\bibliography{references}

\end{document}